## Title

Effect of discrete impurities on electron transport in ultra-short MOSFET using 3D Monte Carlo simulation

## Authors

Philippe Dollfus, Arnaud Bournel, Sylvie Galdin,

Sylvain Barraud[(*)], and Patrice Hesto

Institut d'Electronique Fondamentale
UMR 8622 CNRS / Université Paris-Sud XI
91405 Orsay cedex, France
e-mail: philippe.dollfus@ief.u-psud.fr
[(*)] present address: LETI/DTS, CEA/Grenoble, 17 rue des martyrs
38054 Grenoble cedex 9

## Abstract

This paper discusses the influence of the channel impurity distribution on the transport and the drive current in short-gate MOSFET. In this purpose, a careful description of electron-ion interaction suitable for the case of discrete impurities has been implemented in a 3D particle Monte Carlo simulator. This transport model is applied to the investigation of 50 nm MOSFET operation. The results show that a small change in the number of doping impurities or in the position of a single discrete impurity in the inversion layer may significantly influence the drain current. This effect is not only related to threshold voltage fluctuations but also to variations in transport properties in the inversion layer, especially at high drain voltage. The results are analyzed in terms of local fluctuations of electron velocity and current density. In a set of fifteen simulated devices the drive current $I_{on}$, determined at $V_{GS} = V_{DS} = 0.6$ V, is found to vary in a range of 23% according to the position of channel impurities.

I. INTRODUCTION

In many respects related to physical and technological limitations the downscaling of MOSFETs to sub-50 nm dimensions poses in new terms the design of next CMOS generations. Among the new important issues, the inescapable statistical fluctuations in the number and position of doping impurities yield spreading in device characteristics as the threshold voltage ($V_T$) and the off-state current, to such a point that it may be unacceptable to CMOS operation. This limitation was early predicted by Keyes in 1975 [1] and experimentally demonstrated in the mid-1990s [2-4].

The statistical $V_T$-fluctuations induced by dopant random distribution have been deeply investigated by analytical approaches [3,5] and by numerical drift-diffusion (DD) simulations using either continuous doping [5,6] or 3D atomistic models [7-9]. Even though the DD method is very effective in providing a large amount of results for statistical studies, 3D Monte Carlo (MC) simulations should give a more realistic description of transport, specially in the case of ultra-small device. Moreover, in most of atomistic simulations the discrete nature of impurities is only included in the charge assignment to the grid cells. The description of the long-range part of electron-ion interaction is then accounted for from the solution of the 3D Poisson equation that gives the so-called mesh force acting on the carriers. But it was shown that the short-range part of the interaction must be carefully taken into account and distinguished from the long-range part to correctly describe the Coulomb force [10-15]. 3D MOSFET simulations using a coupled Monte Carlo/molecular dynamics algorithm to properly include the short-range effects were recently reported [16]. In the present article we use a pure Monte Carlo approach which includes the short-range part of the Coulomb interaction via a specific scattering mechanism applied in predefined short-range zones (SRZ). The detailed description of this electron-ion interaction model for 3D particle simulation may be found in [15]. However, some of its most important features are given in the next section to make clear the rest of the article. The model is validated through the computation of electron drift mobility in both N-type and P-type 3D Si resistors and the comparison with experimental results. Sec. III is devoted to the results of 50-nm MOSFET's simulation. We analyze the influence of discrete impurities on the carrier transport in the channel, and the effect of typical small changes in the impurity distribution on the drain current characteristics. This work does not aim at giving a full statistical analysis of $V_T$ fluctuations, but at physically describing the influence of dopant number and position on internal quantities (field, velocity, density) and on the drive current $I_{on}$.

## II. THE ELECTRON-ION SCATTERING MODEL

This section is dedicated to the description of the electron-ion interaction model developed for the case of discrete impurities. It has been shown that the electrostatic field resulting from the solution of 3D Poisson equation cannot correctly describe the interaction in the vicinity of impurities [10-13]. The necessary correction usually consists in defining around each impurity a short-range zone (SRZ) in which the long-range Coulomb interaction is removed and replaced by a short-range mechanism. It has been proposed to couple Monte Carlo (MC) and Molecular Dynamics (MD) techniques to compute the short-range forces between particles inside the SRZs [10,11,13] using the so-called $P^3M$ algorithm [14]. In our MC model the short-range interaction is described by an instantaneous scattering mechanism characterized by a scattering rate [15] as for all other scatterings experienced by carriers. This approach is valid if we take care to consider three questions simultaneously: the choice of the mesh size, the choice of the SRZ size and the model of scattering rate.

The choice of the mesh size is dictated by two considerations. First, one has to save the number of nodes on which is solved the 3D Poisson equation. Second, the mesh size $\Delta X$ must be small enough to correctly describe the smallest space fluctuations of potential to be considered in the simulation. Assuming an equilibrium region with an average concentration of discrete impurities $N$, the smallest space effect to be considered is the variation of Coulomb potential between impurities in order to calculate correctly the long-range carrier-ion interaction. A mesh size of one tenth of the average distance between impurities, i.e. $\Delta X = N^{-1/3}/10$, is proved to be a good compromise [15].

In the case of majority carriers, i.e. attractive impurities, the size of the SRZ must be large enough to avoid inconsistencies between a classical description of particles and a thin impurity potential that looks like a quantum box [15]. But to simplify the calculation of electron-ion scattering rate and to correctly describe the long-range interaction, the SRZ should be as small as possible. Indeed, if the volume of the SRZ is chosen smaller than the volume of the screening sphere one can neglect screening effect in the calculation of scattering rate. Given the choice of mesh spacing $\Delta X$ the good trade-off between these contradictory criteria is to extend the SRZ to two cells around impurities. For instance, if the carrier and impurity concentration is $10^{18}$ cm$^{-3}$, the cells are cubes of side $\Delta X = 1$ nm and the SRZs are cubes of side 5 nm, i.e. containing 125 cells, centered on the impurity cells. It should be mentioned that the choice of the SRZ size is less critical in the case of repulsive impurities (minority carriers) than in the case of attractive impurities (majority carriers).

Since the short-range scattering applies within a volume smaller than the screening sphere, i.e. the sphere of radius equal to the Debye length, we do not have to consider screening effects in the derivation of the scattering rate $1/\tau_{imp}$ [15]. Practically, we calculate $1/\tau_{imp}$ using the Takimoto approach [17] as for the case of continuous doping, with a doping concentration equal to 1 over the volume of SRZ. This model includes the screening effect but we assume a very long screening length $L_0 = 0.4\,\mu m$, which makes the screening influence on the scattering rate negligible. We have checked that by using a standard scattering rate with a screening length deduced from the local carrier density, the calculated mobilities are strongly overestimated.

To validate the model, the electron drift mobility in both N-type and P-type silicon has been computed for comparison with experimental bulk-Si mobilities. The simulated structures are 3D resistors with two terminal contacts. The concentration of discrete doping impurities is in the range $10^{15}$-$10^{19}$ cm$^{-3}$. The number of simulated carriers is adjusted to obtain electrical neutrality in the structures. It should be noted that the number of particles is much higher than the number of ions (typically 2000 times higher), which means that we consider sub-particles [18]. It allows us to reduce the particle noise by statistical enhancement of rare events and to get an average number of particles by cell greater than one, which is crucial to stabilize the solution of Poisson's equation in quasi-equilibrium regime. The 3D Poisson's equation is solved at each time step varying in the range 0.1-1 fs, according to the carrier concentration, i.e. the doping concentration. By applying periodic boundary conditions, i.e. each carrier exiting through a contact is re-injected at the other side with the same wave vector, the number of carriers is constant. To compute the majority electron mobility, N-type resistors are simulated including electrons only. To compute the minority electron mobility we simulate P-type resistors with a given small amount of electrons: if the number of holes required to balance the ion charges is $P_0$ we simulate $1.1 \times P_0$ holes and $0.1 \times P_0$ electrons to maintain the neutrality.

The resulting electron mobility is plotted in Fig. 1 (symbols) as a function of doping concentration and compared to experimental results given in [19] (N-type Si, continuous line) and [20] (P-type Si, dashed line). A very good agreement is found between calculated and experimental mobilities in both cases of majority and minority electrons. It should be noted that this atomistic transport model makes very simply the difference between attractive and repulsive impurities. This difference only results from the 3D solution of Poisson equation, i.e. from the long-range part of the electron-ion interaction.

### III. MOSFET SIMULATION

The atomistic model of impurity interaction has been applied to study the effect of the discrete character of channel dopants in short-gate MOSFET. Our intention is not to perform a full statistical analysis of dopant fluctuations, but to evaluate the influence of some specific impurity positions in the inversion layer on electron transport and device characteristics.

The basic structure of simulated devices is shown in Fig. 2. The channel length $L$ and width $W$ are 50 nm. The junction depth $Z_J$ is 15 nm and the gate oxide thickness $t_{ox}$ equals 1 nm. To generate the reference device the doping implantation steps have been simulated using a Monte Carlo model [21]. In N-type source and drain regions the simulated average distribution of arsenic is approximated by a continuous doping profile decreasing from $5 \times 10^{19} cm^{-3}$ at the surface to $10^{18} cm^{-3}$ at the junction depth. The distribution of dopants is considered as a discrete distribution only in the channel, i.e. in the so-called "active region" of volume $L \times W \times Z_J$. The boron implantation steps were optimized to give a uniform average concentration of $10^{18} cm^{-3}$ on a large depth. It yields about 37 discrete impurities in the active region. At depth higher than $Z_J$ the boron distribution is considered as a continuous concentration approximated by a smooth function. This article deals with the effect of small changes in the distribution of some of the 37 discrete impurities within the active region and more precisely in the inversion layer, which led us to study fourteen additional devices to be compared to the reference. Unless otherwise indicated, all results shown below have been obtained for a drain voltage $V_{DS}$ of 0.6 V.

For a doping concentration of $10^{18} cm^{-3}$ the average distance between impurities is 10 nm. According to the criterion defined above the mesh spacing $\Delta X$ is thus chosen equal to 1 nm. Around each impurity the SRZ is then a cube of side 5 nm. In principle, the size of each SRZ should depend on the local carrier density which may strongly change according to the applied bias and the impurity position. However, if the SRZ size is critical in the case of attractive impurities, it is much less in the case of repulsive ions as in a MOSFET channel. We have thus considered the same constant size for all SRZs.

The 3D Poisson's equation is solved at each time step equal to 0.1 fs with standard boundary conditions [22]. The number of simulated particles, which are still sub-particles, is typically 90000. In addition to electron-impurity interaction, the scattering mechanisms included in the Monte Carlo algorithm are phonon scattering and surface roughness scattering. The acoustic intra-valley phonon scattering is treated as an elastic process and the intervalley phonon transitions, consisting of three *f*-type and three *g*-type processes, are considered via either zeroth-order or first-order transition matrix [23] in agreement with selection rules. The phonon coupling constants given in [22] are used. Surface roughness scattering is treated with an empirical combination of diffusive and specular reflection which correctly reproduces the experimental universal mobility curve [24-25]. A fraction of 14 % of diffusive reflections has been considered in this work.

## A. Reference device

As an introduction to the effects of discrete doping we start with some results obtained for the reference device. The cartography (top view) of the longitudinal current density $J_x$ in the first sheet of cells under the gate oxide is shown in Fig. 3 for $V_{GS} = 0.45$ V and $V_{DS} = 0.6$ V. The presence of three impurities in the inversion layer is indicated by white circles and is characterized by low-current regions around impurities (dark) and high-current regions between them (bright).

To analyze the transport in the channel we define in Fig. 3 two typical lines between source and drain. At $Y = 38.5$ nm the solid line does not go through any impurity and at $Y = 8.5$ nm the dashed line goes through one impurity (at $X_I = 23.5$ nm). Along these two lines the profiles of longitudinal electric field ($E_x$ component), electron density and electron velocity ($v_x$ component) are plotted in Figs. 4a, 4b and 4c, respectively. On the electric field profile plotted in dashed line, the presence of an impurity is characterized by a plateau in the 5 nm-wide SRZ, resulting from the vanishing of the mesh force, and by a repulsive field observed on both sides of the SRZ, i.e. positive on the left and negative on the right.

The presence of the impurity strongly affects the local density by generating an electron accumulation on the left side and a depletion on the right side (Fig. 4b). Moreover electrons considerably slow down around the impurity (Fig. 4c), which explains the weak current density observed locally in the cartography (Fig. 3). The largest part of the current circulates in the regions free of impurity where the electron velocity is higher.

Seeing these strong effects of the presence of an impurity on the local quantities, a significant influence of the number and the position of discrete dopants in the inversion layer can be expected. The rest of the article tackles this problem.

*B. Effect of the position along the channel of a single impurity*

In this sub-section, we analyze the effect of moving one of the discrete impurity in the active region along the source-drain $X$ axis. We consider the impurity, visible in Fig. 3, whose initial position is $X_I = 9.5$ nm, $Y_I = 26.5$ nm. Its position $X_I$ is moved from 2.5 nm to 47.5 nm. We first compare two cases: $X_I = 9.5$ nm (initial case, i.e. reference device) and $X_I = 43.5$ nm.

The cartography of current density is shown in Fig. 5 for $X_I = 43.5$ nm, i.e. in the case where the impurity is near the drain-end of the channel. It is thus in the high-field region (as shown in dotted line in Fig. 6a) and when compared to the previous case of Fig. 3 ($X_I = 9.5$ nm) the current density is now higher and more homogeneous in the source-end region where electrons are injected (Fig. 5).

The electric field and velocity profiles are plotted in Figs. 6 in three cases: along the line $Y = 38.5$ nm without any impurity (solid line, as in Fig. 4b), along the line $Y_I = 26.5$ nm with the impurity in $X_I = 9.5$ nm (dashed line) and along the line $Y_I = 26.5$ nm with the impurity in $X_I = 43.5$ nm (dotted line). If the impurity is in the source-end of the channel (dashed line) the injection-velocity is strongly degraded when compared to the case without impurity, but the overshoot velocity at the drain-end is similar in both cases (Fig. 6b). On the contrary, if the impurity is in the drain-end (dotted line) the maximum velocity significantly decreases but the injection-velocity is not affected when compared to the solid line.

The consequence on the drain current is shown in Fig. 7a where $I_D$ is plotted as a function of the gate voltage $V_{GS}$ (at $V_{DS} = 0.6$ V) in both cases. The current is about 3.8% higher for $X_I = 43.5$ nm (dashed line) than for $X_I = 9.5$ nm (solid line). The degradation of the injection velocity at the source-end is thus more detrimental than the reduction of overshoot velocity at the drain-end, which is in agreement with other theoretical investigations [26-27]. Additionally, it should be noted that the difference between both curves is not a shift but rather a change in the slope, i.e. in the transconductance. In other words, changing the $X$ position of an impurity induces a change in the channel transport properties which reflects on the drive current but probably not on the threshold voltage $V_T$. It is confirmed by the $I_D$-$V_{DS}$ characteristics plotted in Fig. 7b. At low $V_{DS}$, i.e. under the usual conditions of $V_T$ measurement, the current results from the average transport properties all along the effective channel length and is thus very weakly influenced by the *X*-position of a single impurity. Differences between these two devices only occur at high $V_{DS}$, that is when the current is strongly dependent on the injection velocity at the source-end and thus influenced by the impurity position.

The drive current $I_{on}$, defined at $V_{GS} = V_{DS} = 0.6$ V, is plotted in Fig. 8 as a function of the impurity position $X_I$ (solid line linking the full circles). The general trend to the increase of $I_{on}$ as the position $X_I$ goes to the drain is confirmed. However, the current slightly increases also when the impurity goes very close to the source-end. In this case, indeed, the SRZ associated with the impurity and the repulsive field on the left side partially extend in the highly-doped source region. The influence of this impurity on the electron transport is then smaller. To be fully effective the ion must be plainly inside the active region, i.e. at $X_I \geq 10$ nm.

## C. Effect of an added or removed single impurity

We now analyze the effect of changing the number $N_{imp}$ of impurities in the inversion layer by one unit. There are 37 impurities in the active region of the reference device. We already mentioned the effect of the impurity located at position $X_I = 23.5$ nm and $Y_I = 8.5$ nm (sub-section III.A), i.e. at a *X* position for which the effect of the impurity on the total current is maximum, as just shown (in Fig. 8). We consider two other devices by removing this impurity, yielding $N_{imp} = 36$, or by adding an impurity at the same *X* position but a different *Y* position ($Y_I = 38.5$ nm), yielding $N_{imp} = 38$. The cartography of current density in these two devices is shown in Figs. 9a and 9b, respectively, to be compared with Fig. 3. As expected, the homogeneity of the current density in the source-end region is obviously strongly dependent on these configurations, which should have significant effects on the total drain current. It is shown in Fig. 10 where $I_D$ is plotted as a function of $V_{GS}$ for the three devices. In this case the curves are almost parallel: the most important effect is thus a $V_T$ shift. Let come back to the Fig. 8 where the $I_{on}$ value of the two new devices is plotted for comparison (open diamond and open square). The influence of the presence of a single impurity on $I_{on}$ appears to be about 3.3%.

Of course, in the case $N_{imp} = 36$, the effect on the current depends on the initial position of the removed impurity because of the variable electrostatic influence of other surrounding impurities. As shown in Fig. 8 for this specific simulated device, removing the impurity located at $X_I = 23.5$ nm (open diamond) yields a slightly smaller $I_{on}$ value than removing the impurity located at $X_I = 9.5$ nm (closed diamond). It should be noted that removing the impurity or placing it in the drain-end region induce very similar effects resulting in very close values of $I_{on}$.

The effect of trapping of a single electron has been recently studied using atomistic DD simulation [28]. It is interesting to remark that relative current fluctuations of 5 % have been observed in strong inversion (at low drain bias), which is quite comparable with the effect of a single impurity obtained in the present work (at high drain bias).

## D. Effect of the vertical position of some impurities

We now consider the effect of moving impurities vertically. Indeed, at a given number of impurities in the full active region ($N_{imp} = 37$ in our case), changing the number of impurities in the inversion layer is expected to modify the overall transport properties in the channel and the device performance. Starting from the reference device, another transistor has been generated by burying the three impurities visible just under the gate oxide (Fig. 3) 3.5 nm below the surface. This gives the current density mapping shown in Fig. 11a. The current density is then much more homogeneous in the inversion layer, even if buried impurities still have a small influence.

On the contrary, the cartography shown in Fig. 11b corresponds to a device where four buried impurities have been moved to the surface, which yields the presence of seven impurities in the inversion layer. Two of these new surface impurities are close to the source-end and the two other ones are near the drain. Thus, as shown in sub-section III.B, they are not in a position where they can have the maximum influence on the transport. To enhance the effect of these four impurities they have been distributed along a line $X_I = 11.5$ nm, which results in the cartography displayed in Fig. 11c. The regions of high current density are now very limited when compared to the previous distribution shown in Fig. 11b and *a fortiori* to that of the reference device (Fig. 3).

The consequence on the current characteristics and the value of $I_{on}$ appears clearly in Fig. 12. Compared with the reference device (solid line), at $V_{DS} = 0.6$ V and $V_{GS} = 0.45$ V the current increases by about 9% by burying the three surface impurities (squares, dotted line) and decreases in the worst case by 22% with seven surface impurities (diamonds, dashed line).

At $V_{GS} = 0.6$ V these figures become 5.8% and 16.8%, respectively. These current changes seem to be the consequence of both transconductance and threshold voltage fluctuations.

## *E. Summary - Current statistics*

The number of simulated devices is certainly too small to make an accurate statistics on the effects of random fluctuations of doping impurity distribution in the inversion layer. However, we plot in Fig. 13 the bar chart of $I_{on}$ values obtained for the fifteen devices, which summarizes the results obtained. This figure includes the result obtained by reversing source and drain in the reference device which increases $I_{on}$ by 2.9%.

In this set of devices the average $I_{on}$ value is $\langle I_{on} \rangle = 30.9$ µA. The maximum and minimum $I_{on}$ values are 32.7 µA and 25.6 µA, respectively. The simple effect of changing the position of a few impurities or adding/removing a single impurity in the inversion layer may thus induce drive current fluctuations in a range of at least 23% for this 50 nm MOSFET architecture. If we eliminate the worst case which is unlikely ($I_{on} = 25.6$ µA, corresponding to the distribution of Fig. 11c with six impurities almost aligned in the source-end region) the range of $I_{on}$ fluctuations is still 11.2%. The relevant figure is probably in between these two values and thus is not negligible. We would like to point out again that these random current fluctuations observed at high drain voltage are not only due to threshold voltage shift but also in some cases to changes in channel transport properties reflected on the transconductance. A possible extension of this work could be the investigation of donor fluctuations, especially in the source region.

## IV. CONCLUSION

Within 3D Monte Carlo simulation we have developed an electron-ion interaction model suitable for investigating the effect of dopant random fluctuations. The effect of fifteen typical distributions of channel doping atoms has been analyzed for a 50 nm MOSFET in terms of local physical quantities (as electron density and velocity, current density) and drain current. Small variations in the number and the position of these atoms are shown to significantly influence the transport properties in the inversion layer, which results not only in threshold voltage fluctuations but also, at high drain voltage, in transconductance variations. As a consequence, the values of drive current are spread out on a range higher than 11% and likely to reach more than 20%. It should be still larger in more aggressively downscaled devices, which probably reinforces the interest in investigating MOSFET architectures with undoped thin-channel for future CMOS generations.

# FIGURE CAPTIONS

Figure 1. Electron mobility as a function of discrete dopant concentration in N-type and P-type Si resistors. Solid and dashed lines are fitting curves of experimental results given in [16] (N-type) and [17] (P-type). Symbols are our simulation results.

Figure 2. Schematic view of simulated 3D MOSFETs with discrete impurities in the active region.

Figure 3. Cartography of current density $J_x$ in the first sheet of cells under the gate oxide (top view). $X = 0$ and $X = 50$ nm correspond to the source/channel and channel/drain N-P junctions, respectively. The position of the three impurities present in the inversion layer is indicated by white circles (reference device). The solid and dashed lines indicate the lines along which the curves of Fig. 4 are plotted.

Figure 4. Profiles of electric field $E_x$ (a), electron concentration (b) and electron velocity $v_x$ (c) between source and drain along two lines: along $Y = 38.5$ nm (solid line), i.e. without impurity; and along $Y = 8.5$ nm (dashed line) with an impurity in $X_I = 23.5$ nm (see Fig. 3).

Figure 5. Cartography of current density $J_x$ in the device where the impurity initially in $X_I = 9.5$ nm (reference, Fig. 3) has been moved to $X_I = 43.5$ nm.

Figure 6. Profiles of electric field $E_x$ (a) and electron velocity $v_x$ (b) along three lines: along $Y = 38.5$ nm without any impurity (solid line) and along $Y = 26.5$ nm with one impurity either in $X_I = 9.5$ nm (dashed line) or in $X_I = 43.5$ nm (dotted line).

Figure 7. (a) Drain current versus gate voltage for two devices differing from one another in the X position of one impurity ($V_{DS} = 0.6$ V) ; (b) Drain current versus drain voltage for the same devices ($V_{GS} = 0.6$ V).

Figure 8. Drive current $I_{on}$ as a function of the position of one impurity in the channel

between source and drain along the line $Y = 26.5$ nm ($N_{imp} = 37$, circles). The graph also shows the results obtained by removing this impurity ($N_{imp} = 36$, closed diamond) and by adding ($N_{imp} = 38$, open square) or removing ($N_{imp} = 36$, open diamond) another impurity at $X_I = 23.5$ nm.

Figure 9. Cartography of current density $J_x$ in two devices where one impurity has been removed (a) or added (b) at $X_I = 23.5$ nm (to be compared with the reference, Fig. 3).

Figure 10. Drain current versus gate voltage for three devices differing from one another in the number of impurities in the version layer: the three curves correspond to the devices whose current cartography is shown in Fig. 3 (solid line), Fig. 9a (dotted line) and Fig. 9b (dashed line) ($V_{DS} = 0.6$ V).

Figure 11. Cartography of current density $J_x$ in three devices differing from one another in the vertical position of surface impurities (with $N_{imp} = 37$). When compared to the reference (Fig. 3), three surface impurities are buried 3.5 nm below the surface (a) or four buried impurities are moved to the surface (b and c) at different positions.

Figure 12. Drain current versus gate voltage (at $V_{DS} = 0.6$ V) for four devices differing from one another in the vertical position of some impurities, i.e. in the number of surface impurities which is either 3 (reference, solid line), 0 (dotted line) or 7 (dashed lines). These devices correspond to the current cartographies of Figs. 3 (circles), 11a (squares), 11b (triangles) and 11c (diamonds).

Figure 13. Bar chart of $I_{on}$ values of all simulated devices. The dashed line indicates the average value $\langle I_{on} \rangle$.

Dollfus et al.  **Figure 1**

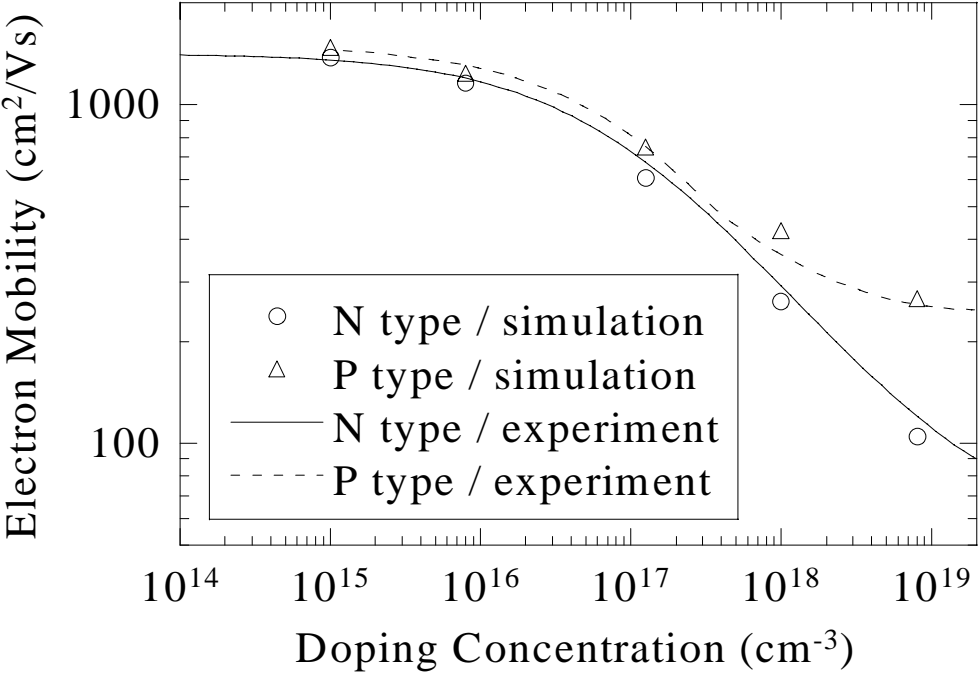

Dollfus et al.    **Figure 2**

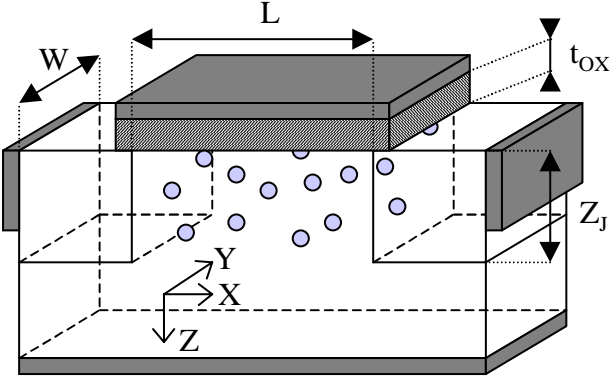

Dollfus et al.  **Figure 3**

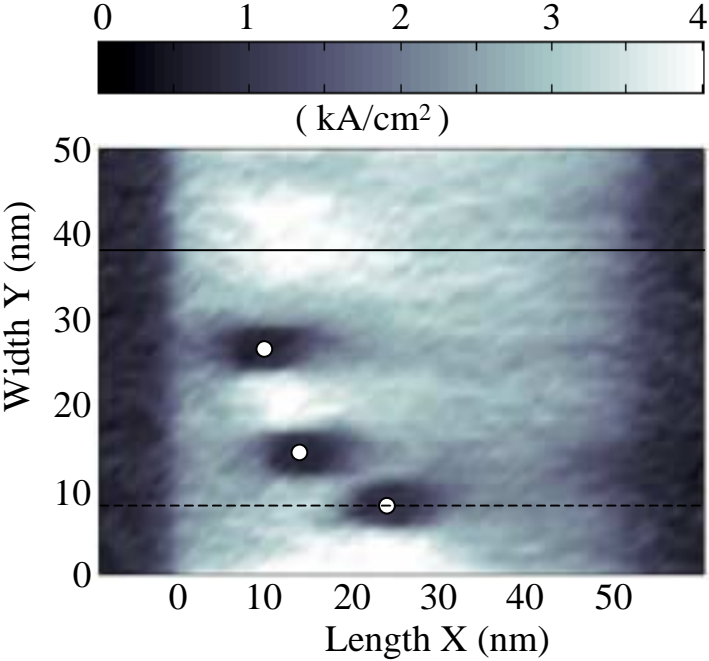

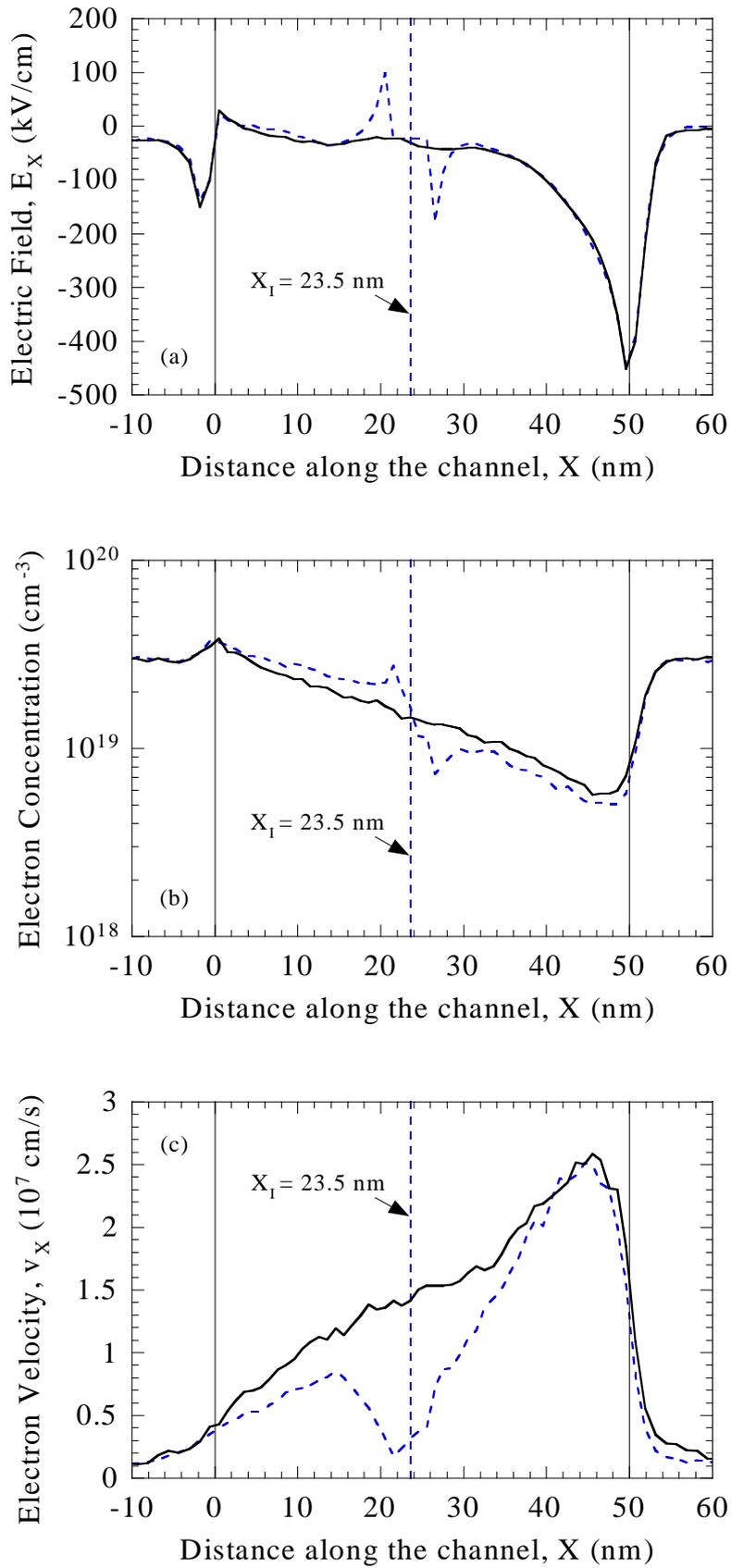

Dollfus et al.    **Figure 5**
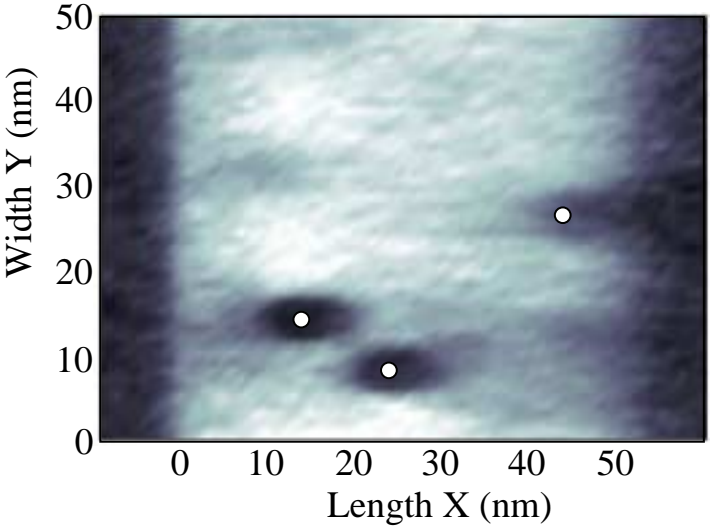

Dollfus et al.  **Figure 6**

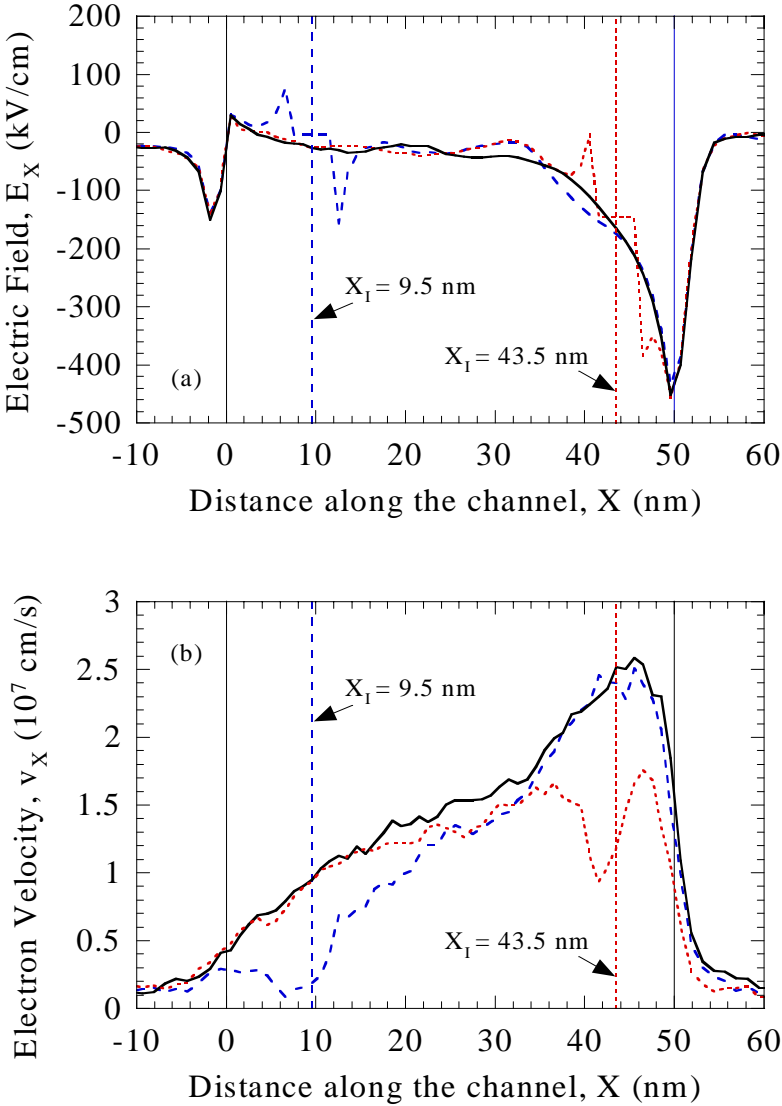

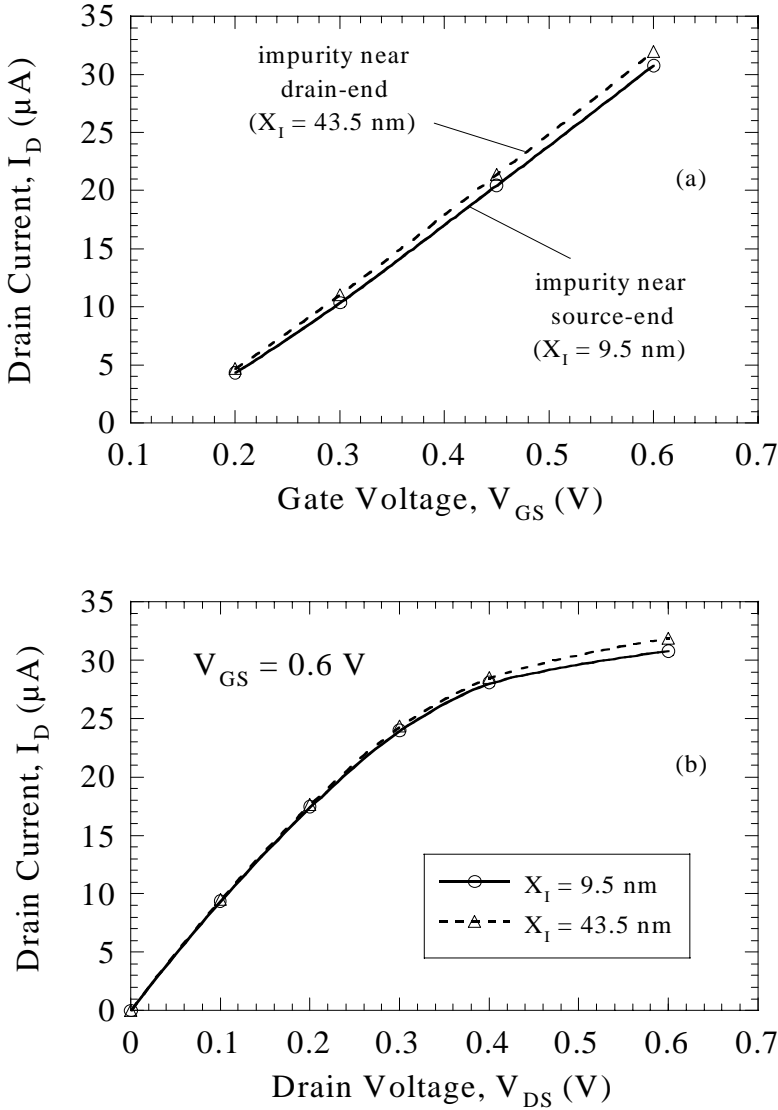

Dollfus et al.     **Figure 8**

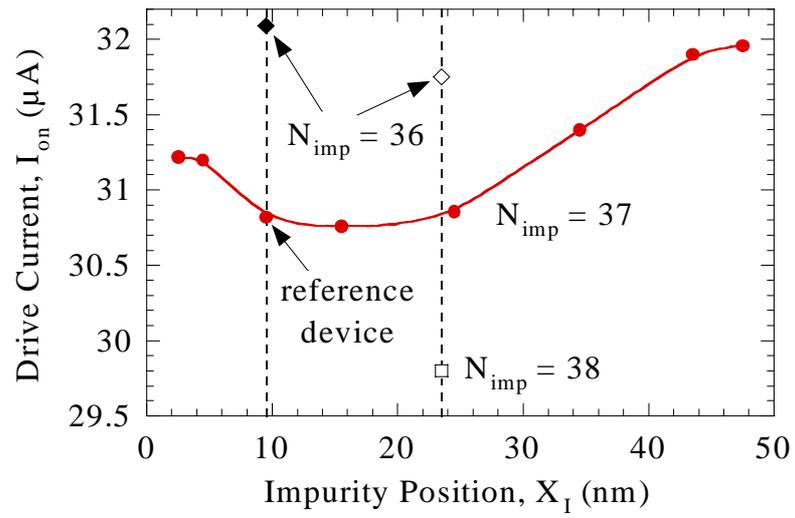

Dollfus et al. **Figure 9**

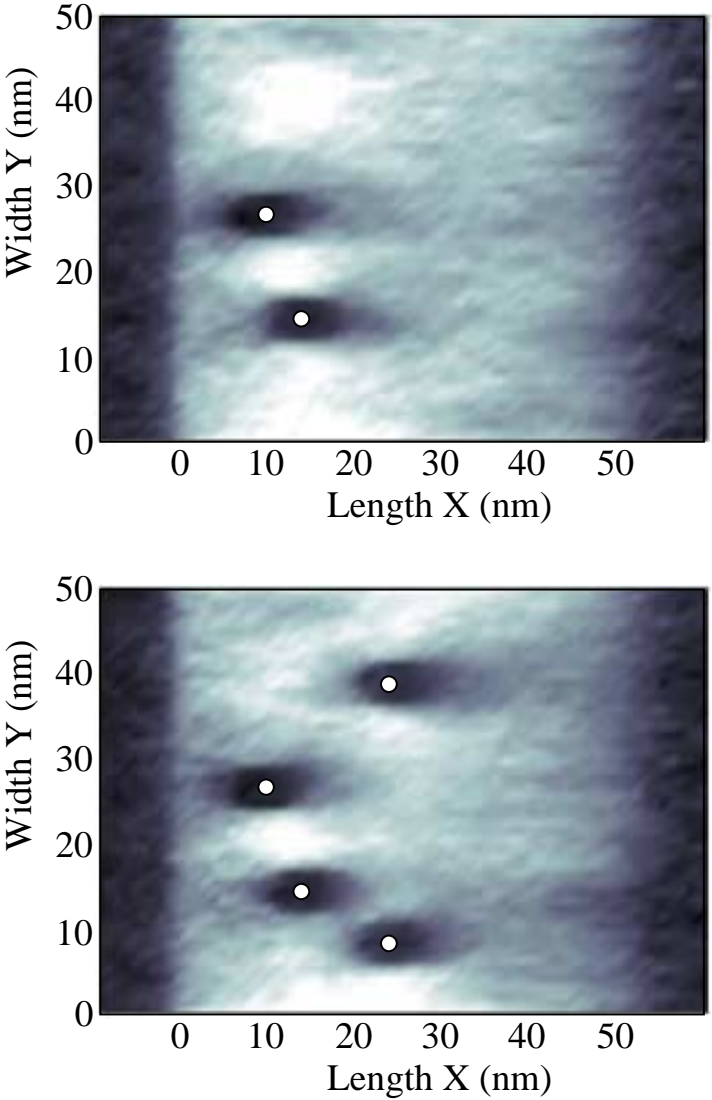

Dollfus et al.  **Figure 10**

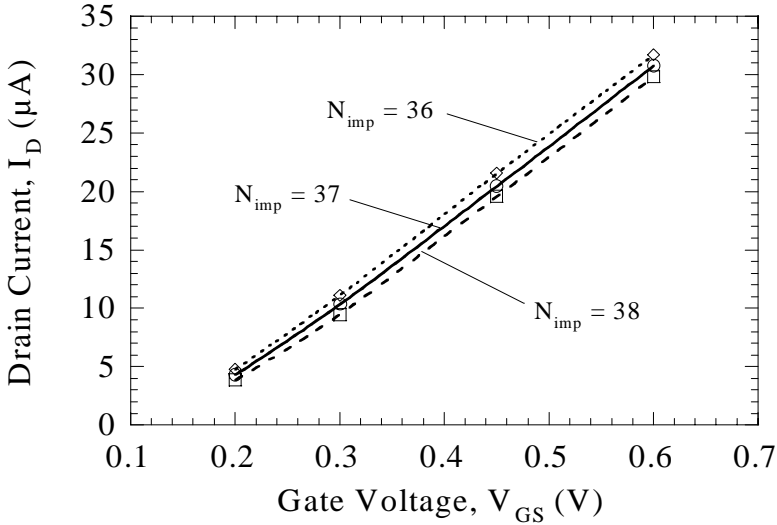

Dollfus et al.  **Figure 11**

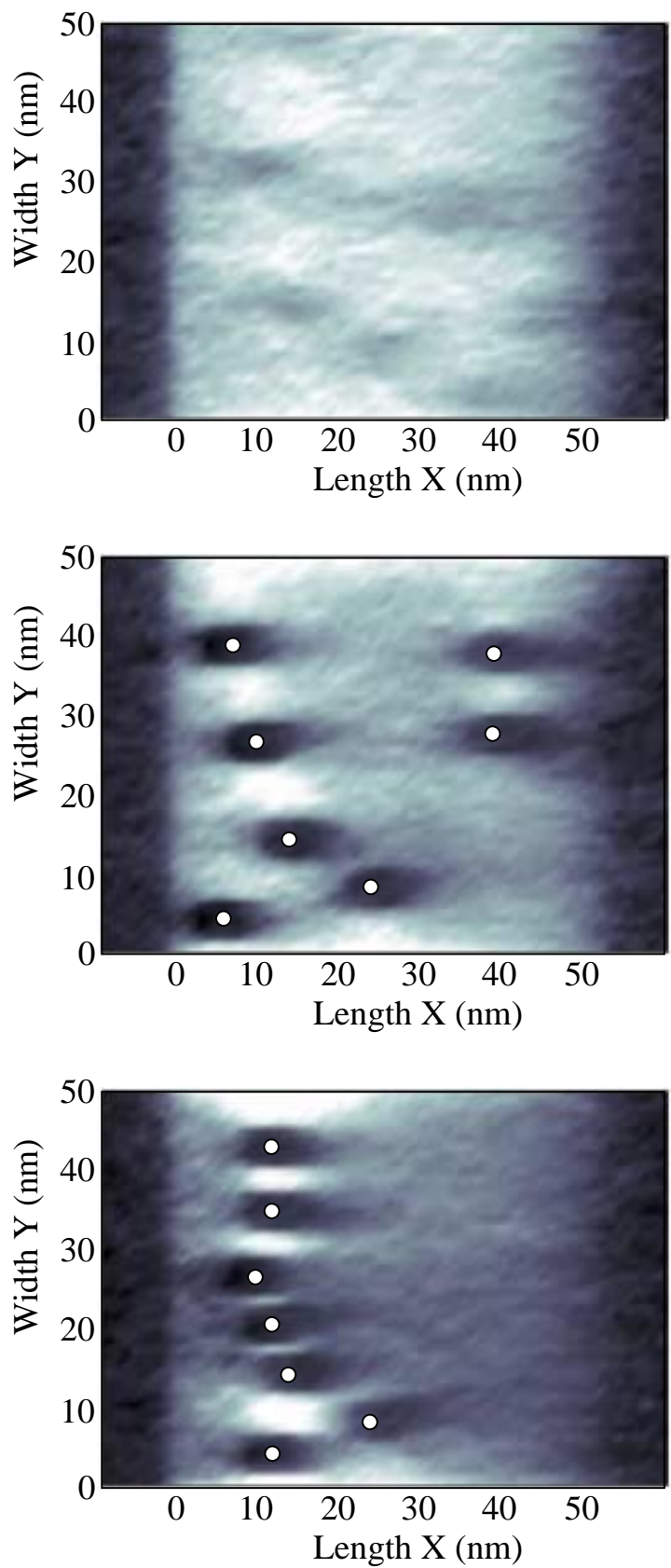

Dollfus et al.  **Figure 11**

Dollfus et al.   **Figure 12**

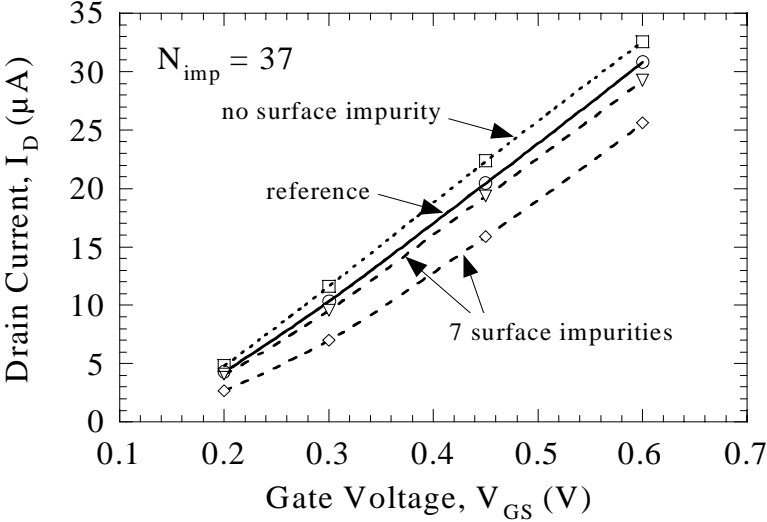

Dollfus et al. **Figure 13**

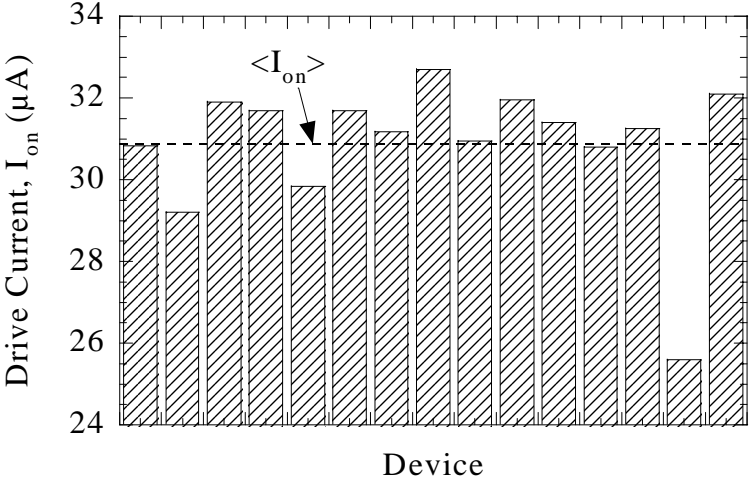